# SHORT-PERIOD WAVES THAT HEAT THE CORONA DETECTED AT THE 1999 ECLIPSE

JAY M. PASACHOFF[1], BRYCE A. BABCOCK, KEVIN D. RUSSELL, and DANIEL B. SEATON[1]

Hopkins Observatory, Williams College, Williamstown, Massachusetts 01267, U.S.A.

**Abstract.** As a part of a study of the cause of solar coronal heating, we searched for high-frequency (~1 Hz) intensity oscillations in coronal loops in the [Fe XIV] coronal green line. We summarize results from observations made at the 11 August 1999 total solar eclipse from Râmnicu-Vâlcea, Romania, through clear skies. We discuss the image reduction and analysis through two simultaneous series of coronal CCD images digitized at 10 Hz for a total time of about 140 s. One series of images was taken through a 3.6 Å filter isolating the 5303 Å [Fe XIV] coronal green line and the other through a 100 Å filter in the nearby K-corona continuum. Previous observations, described in Pasachoff *et al.* (2000), showed no evidence for oscillations in the [Fe XIV] green line at a level great than 2% of coronal intensity. We describe several improvements made over the 1998 eclipse that led to increased image clarity and sensitivity. The corona was brighter in 1999 with the solar maximum, further improving the data. We use Fourier analysis to search in the [Fe XIV] channel for intensity oscillations in loops at the base of the corona. Such oscillations in the 1-Hz range are predicted as a result of density fluctuations from the resonant absorption of MHD waves. The dissipation of a significant amount of mechanical energy from the photosphere into the corona through this mechanism could provide sufficient energy to hear the corona. A Monte-Carlo model of the data suggests the presence of enhanced power, particularly in the 0.75-1.0 Hz range, and we conclude that MHD waves remain a viable method for coronal heating.

## 1. Introduction

During the last few decades a wide array of theories of the mechanism that heats the solar corona to temperatures above one million degrees have been propagated through the solar physics community. It is widely accepted that energy is provided by some combination of processes (Golub and Pasachoff, 1997, 2001; Priest and Forbes, 1999; Roberts, 2000; Aschwanden *et al.*, 2001). In quiet regions, microflares or nanoflares—small impulsive events—have been postulated to play an important role in moving energy into the corona (Krucker and Benz, 2000; Krucker *et al.*, 1997, Gomez *et al.*, 2000). In active regions, transient x-ray and EUV bursts have been observed by SOHO, Yohkoh SXT, and TRACE that may play a role in heating (Shimizu *et al.*, 1994; Berghmans *et al.*, 2001). While space based observation is well suited for the detection of events in x-rays and the EUV, only ground based observation during a total solar eclipse provides a high-resolution, low-scattering view of the lowest corona in visible and near ultraviolet wavelengths.

Following our search for short-period coronal oscillations at the 1994 and 1998 total solar eclipses (Pasachoff *et al.*, 2000), we sought to improve our detection capability significantly at the 1999 eclipse. As discussed in our previous paper, our experiment was first designed to test the surface-Alfvén-wave model of coronal heating (Ionson, 1978, 1979) and now is also relevant to other MHD models (Hollweg 1981, 1991; Porter, Klimchuk, and Sturrock, 1994a,b; Ofman, Davila, and Shimizu, 1996, Priest and Forbes, 1999). Models by Porter *et al.* (1994a,b) show that waves with periods between 0.1 s and 1 s could play a significant role in the heating of the corona. These fast-mode MHD waves on the level of 1% of total power could carry energy to heat the corona to millions of kelvins. To dissipate that energy as heat, they are necessarily damped, and hence only quasiperiodic. Since our detection capabilities at the 1998 eclipse were approximately 2% for a sustained oscillation, an improved experiment was necessary to test the theory conclusively, especially since our narrow-band filter may have been only partially on-band in 1998.

---

[1] Now at Harvard-Smithsonian Center for Astrophysics, 60 Garden St., Cambridge, MA 02138, USA.

Our experiment looks for oscillations in both the continuum near 5570 Å and the [Fe XIV] line at 5303 Å. The continuum observations, which detect oscillations proportional only to the density of the coronal loop, provide a check on intensity fluctuations that might result from instrumental error or atmospheric effects. That is, the ratio of the iron line images to the continuum images should reveal only real coronal oscillations. Our spatial resolution is 2.2 arcsec, which is about 1500 km on the sun, and our temporal resolution is determined by our image acquisition rate of 10 Hz.

Motivated by previous findings by Pasachoff and Landman (1984) and Pasachoff and Ladd (1987), Singh *et al.* (1997) searched for short-period white light oscillations at the 24 October 1995 eclipse using a cooled photomultiplier at an acquisition rate of 20 Hz, an aperture that subtended 90 arcsec at 1.25 solar radii, and a 240 Å continuum passband around 5500 Å. They report significant oscillations of 0.2% to 1.3% of total coronal intensity at 6 frequencies, all within the 0.01 to 0.2 Hz range.

The same group observed with a slightly modified setup at the 26 February 1998 eclipse. Cowsik *et al.* (1999) report that white light observations were carried out with 6 photomultipliers subtending 18 arcsec, at locations of 1.2 and 2 solar radii. Data were taken at 20 Hz in three channels and 50 Hz in three channels. They used 150 Å passbands near 4700, 4900 and 5000 Å. They report significant oscillations of 0.5% to 3.5% of total coronal intensity at 3 frequencies between 0.01 Hz and 0.15 Hz. Even 18 arcsec is several times the width of a coronal loop, thus is it unclear how oscillations could be in phase over their observation area despite their experimental checks ruling out other signal sources, as discussed in Russell (2000). We, therefore, prefer our higher spatial resolution and narrow band filter to their white-light experiment.

Phillips *et al.* (2000) and Williams *et al.* (2001) report successful CCD observations of the 11 August 1999 eclipse, using a two camera setup in an experiment modeled after ours. They observed at close to 50 Hz with 12 bit resolution. Our slower 10 Hz resolution allows us to image at 16 bit dynamic range and therefore distinguish more levels of coronal intensity by a factor of 24. The initial analysis of their data showed the presence of 6 second oscillations with more than 99% significance, but was inconclusive for higher frequencies. They obtained further data in Zambia at the total eclipse of 21 June 2001.

All these experiments use total solar eclipse observations to look for high-frequency waves because none of the spacecraft that observe the sun are currently capable of observing within this high-frequency domain, and eclipses are the only time that background light levels are low enough to observe the corona from Earth with sufficiently low noise, a condition not obtained by mountaintop coronagraphs. The different spatial and temporal resolutions of the experiments provide good complementary analysis.

## 2. Observations

We observed from the roof of the Alutus Hotel in Râmnicu-Vâlcea, Romania, on 11 August 1999 with clear skies and several improvements in our apparatus from the version used at the 1998 eclipse. As at the 1998 eclipse, we used a Princeton Instruments TE/CCD 576FTE camera, with a 36-cm Celestron Schmidt-Cassegrain telescope and an image again split into two channels. The 'green-line channel' contained a 3-Å FWHM filter centered on 5303 Å [Fe XIV], and the 'continuum channel' contained a 100-Å FWHM filter centered near 5570 Å, chosen to avoid chromospheric and coronal emission lines. Each image of our data sequence was taken using our CCD detector with a Princeton Instruments ST-138 controller. In order to take images at 10 Hz, the CCD was run in frame transfer mode and read-out binned 2×2, giving us a 192×144 data-pixel frame. This equipment was run using IPLab software on a Power Macintosh 8500/120 with 200 MB of RAM. Our field of view was 310×200 arcsec for each channel (half of the 192×144 frame, or roughly 96×144), so our pixel scale is 2.16 arcsec pixel$^{-1}$. In addition to allowing a faster readout, our choice of 2×2 binning for our CCD readout has the effect of increasing signal-to-noise for each data-pixel and reducing CCD pixel cross-talk.

Two important modifications were made to the apparatus between 1998 and 1999 that significantly improved the quality of observations. The first of these improvements was the introduction of an OceanOptics S2000 Fiber Optic Spectrometer. The need for such an addition came from the suspicion that the filter in the line-channel was not precisely centered at 5303 Å during the 1998 eclipse, when there was no way to do an on-site empirical check, because of a lower than expected ratio of line signal to continuum signal. Because the centering of the passband is temperature dependent, the spectrometer allows us to adjust the temperature of the filter until we are certain it is centered on-band. A fiber optic pick-off that could be lowered into the optical path of the line-channel was added to the optics box. The pick-off could then be raised out of the optical path during data acquisition.

The second improvement to the optical apparatus was the substitution of a Losmandy equatorial mount for the Byers mount that was used at the 1998 eclipse. Though frames were aligned in data processing to better than 1 pixel, it is best to limit the mount of drift that has to be removed. The Byers mount had a periodic error with a period of 140 seconds of time and an amplitude of 6 arcsec. Therefore, in the most extreme case, a coronal feature could drift as much as 12 arcsec, or approximately 6 binned pixels in our image, during a 700 frame eclipse sequence. Given the inherent limitations on our alignment, mentioned below, and the problematic noise that long-term drift introduces as a result at low frequencies, it was beneficial to attain a mount that reduced this effect. On-site tests indicate that the periodic error for the Losmandy mount had a period of 240 seconds, with an amplitude of approximately 3 arcsec. This means that the worst case scenario for a 700 frame sequence would be approximately 6 to 7 arcsec of drift, or 3 to 4 binned pixels, because the 70 seconds of such a sequence is just over a quarter of the period of the periodic error. While this is not a huge improvement, it reduces drift to within the same magnitude as jitter and atmospheric effects.

In addition to improved drives, the Losmandy mount offered two other advantages. First, it includes a very sturdy portable base structure, aiding in expeditionary set-up. Second, the Losmandy included a paddle control to slew the telescope. This not only made alignment on any object considerably easier, but also, more importantly for us, it allowed the ability to switch the pointing of the telescope from one limb of the moon to the other quickly and accurately. The limb switch was necessary during the 1999 eclipse because, although we had planned to observe of the West limb of the sun because of the presence of observable coronal loops there, orientation during the last few seconds before totality was only possible using the remaining sliver of the sun on the East limb. This switch had been performed manually with the Byers mount, which caused a recoil from which it took several seconds to recover, a significant time during an eclipse totality.

In order to be sure that we made our observations in a region of the sun with strong loop structure, we received space-based support from the *Transition Region and Coronal Explorer (TRACE)*. The best candidates for finding coronal loop oscillations are the bright low parts of loops associated with active regions. TRACE provided EUV images revealing high temperature (log $T \approx 6$) loops with spatial resolution of 0.5 arcsec. Figure 1 shows a TRACE image taken in the 195 Å passband (Fe XII) taken a few hours before the our observations were made. Using our web access and an Inmarsat satellite phone, we agreed with TRACE coordinators to point at 70.2° West of solar North. Further support came from the *Solar and Heliospheric Observatory's (SOHO) Large Angle Spectroscopic Coronal Observatory's (LASCO)* C2 and C3 coronagraphs, which provided images of the extended corona from 2.5 solar-radii to 30 solar-radii and the *Extreme-Ultraviolet Imaging Telescope (EIT)*, which provides full-disk views in EUV wavelengths.

Figure 2 shows a comparison of data frames captured in 1998 and 1999. Even visual inspection reveals finer, stronger features in the line channel. A quantitative evaluation shows the increase in data quality in the ratio of counts in the line channel to counts in the continuum channel. This ratio varies over the image, increased by a factor of four from the 1998 to the 1999 experiment. We therefore conclude that our data quality is indeed better, most likely because the spectrometer monitoring insured that the line filter was on band.

Further increase in data quality was the result of improved weather conditions. While the 1994 and 1998 expeditions had to deal with clouds and strong winds, during the eclipse the skies were entirely clear and wind was essentially non-existent. This reduced noise both as a result of telescope vibration and reduced atmospheric transmission. Further, the corona itself was closer to solar maximum and substantially brighter in 1999 than in 1998, not to mention brighter than the minimum corona of 1994.

## 3. Data Reduction and Analysis

Each image series was processed on an Apple Power Macintosh 8500/120 with 200 MB of RAM (later replaced by an Apple Power Macintosh G4 with 512 MB of RAM) using the IDL software package of Research Systems, Inc. Processing began with subtracting the combined bias and dark field images and dividing out the flat-field image to correct for instrumental noise. We then aligned the lunar limb in each frame to a chosen reference frame using cross-correlation techniques to correct for errors in the tracking of the telescope (see Figure 3). Following the correction of the tracking errors, the lunar limb in each frame was offset in increments of 0.0168 pixels per frame in the geocentric east-west direction and 0.00517 pixels per frame north-south. This process restored the proper lunar motion with respect to the sun, according to calculations made for us by Fred Espenak, NASA Goddard Space Flight Center. This adjustment gave a time series in which the position of the solar corona was constant in each image, to within the error of the alignment, approximately 1 arcsec.

In properly reduced and aligned images the intensities of both the line and continuum signals depend on various factors:

$$I(\lambda) = [K(\lambda) + F(\lambda) + S(\lambda) + E(\lambda)] T_A(\lambda) T_F(\lambda), \qquad (1)$$

where $K$, $F$, $S$, and $E$ are the contributions from coronal electron scattering, scattering from interplanetary dust, atmospheric scattering, and emission lines, respectively. $T_A$ and $T_F$ are atmospheric and filter transmission factors. In order to ensure that fluctuations measured are those in $E(5303\ \text{Å})$ alone, we calculate the ratio of the intensity of the 5303 Å emission line to that of continuum light at 5303 Å ($I_{5303}=I_{C(5303\ \text{Å})}$) at a particular pixel. Because we are so near the peak of the Planck curve, the continuum intensity at 5570 Å is within 5% of the continuum intensity at 5303 Å, so that using ($I_{5303}=I_{C(5570\ \text{Å})}$) in place of ($I_{5303}=I_{C(5303\ \text{Å})}$) merely introduces two negligible errors, described in our previous paper. The flat-fielding procedure makes the calculation straightforward for the following reasons. In addition to removing the effects of inhomogeneities in the detector and optical system, the flat-fielding accounts for the effects of aperture size and filter bandwidth for each image. By flat fielding, we have equalized the number of counts in each channel not resulting from the emission line, so that we can then take the difference between the two channels in order to get the line intensity. Furthermore, this equalization allows us to divide the calculated emission line intensity by the measured intensity of the continuum channel to attain the desired ratio:

$$I_{E(5303)} / I_{C(5303)} = (I_L - I_C)/I_C \qquad (2)$$

where $I_L$ is the measured intensity of the line channel and $I_C$ is that of the continuum channel. Continuum light in the corona comes from light being scattered by coronal electrons, so continuum intensity is proportional to coronal density. We can therefore expect a change in light intensity to accompany a compression wave that changes density. The intensity of light in the [Fe XIV] line is proportional to the square of density, because it is the result of coronal excitation. Since the number of collisions goes as the density squared, so does the intensity of light at the wavelength of the emission line. Therefore, taking this line-to-continuum ratio should separate out variations in intensity of light due to the waves we seek from other variations.

This calculation was used in the analysis of the 1998 data with some success. However, while our alignment within a single channel was good to 1 arcsec, a perfect correction for the offset channel to channel was more difficult to achieve. Data models (described in detail below) for the 1999 data showed that when the line and continuum channel images could not be perfectly aligned this calculation increased background noise levels significantly. Because even only a small amount of misalignment between line and continuum led to a large increase in noise, we decided to focus solely on the line channel.

In order to search for oscillations in the time series of a pixel or group of pixels on a coronal loop, we used the IDL software package to take the discrete Fourier transform (DFT) of each time series, as we did for data sets from previous eclipses. By examining power spectra of these time series, we could determine what frequencies, if any, showed signs of enhanced power. Because it is likely that any impulsively generated short-period waves would contain only quasi-periodic oscillations lasting for a few tens of seconds (Berghmans *et al.*, 1996), we did not perform transforms of the whole image series, which would likely average out any finite-length wavepackets. Instead we used a windowed Fourier analysis, analyzing 100 to 600 frame (10 to 60 second) sequences. We found that a 20 second window provided the best balance between Fourier signal-to-noise and sensitivity to quasi-periodic oscillations. In this paper we focus on transforms of roughly 20 s of data near the end of the eclipse, while signal was highest because we were at the limb where third contact was about to occur, but no saturation had occurred.

In our analysis of the 1998 data, we carried out DFTs on a variety of points and in order to ascertain the significance of a peak in the Fourier spectra, we compared it to the mean power in the spectra. We also compared bright and quiet coronal points to see if peaks occurred more often along the loops. These techniques are described more completely in our previous paper. Since that time, however, we have developed our techniques to allow a more precise approach to the problem, and have concentrated on applying them to the superior data from the 1999 eclipse.

The most significant improvement to the data analysis procedures was the development of comprehensive data models that allowed us to compare our observed power spectra to simulated power spectra—with or without signal present. The first of these models simulates a result with no added signal, which we call a no-effect result. We begin with a realistic counts-per-pixel image generated by averaging a series of 20 images from the data, chosen to strike a balance between maintaining sharply resolved structure and reducing image noise. From this template we generate a series of 200 identical images, to which we add noise with a standard deviation equal to the number of counts in each pixel of the images. Each image is then shifted using offsets measured in the alignment of the actual data sequence. Because skies were nearly photometric during the eclipse, variation of atmospheric transmission is not a significant concern and is not included in the model.

At this point, DFTs of the pre-alignment no-effect model reveal power spectra with nearly identical noise rates to those of the power spectra from pre-alignment data series. This suggests that the largest component of noise is the result of frame-to-frame misalignment. If a bright coronal feature is misaligned by only a fraction of a pixel it can introduce a change in intensity of several percent in any pixels that it spans. We were concerned about this problem because of results we had seen in our analysis of the real data. Figure 4 shows a comparison of the power spectra of two nearby points in the post-alignment real data, one on top of a bright loop and one just off the loop. The on-loop point shows moderately enhanced power at several frequencies, which means that either we are detecting the presence of oscillations or that alignment errors had introduced noise that would mimic coronal oscillations. This analysis raised the question of how positional shifts introduce noise into the system and whether careful alignment removes or reduces the misleading Fourier peaks. (Our later tests using this and similar models, explored in detail below, suggests that, after alignment, high-level noise is not present in bright loops.)

The artificial set was then aligned using similar procedures to those used on real data. DFTs revealed that noise levels in the real data were more variable than those of the artificial set. Further, the mean power of the real data was about 10% higher than the mean power of the model. However, the noise levels are not terribly dissimilar and thus could be used in a 'Monte Carlo' analysis (so named because the noise introduced into the artificial model has a random component) to test the significance of a peak in the spectrum. Several more tests were done using this and other similar models, but before we can discuss them we must introduce the other major improvement to the data analysis package.

As we mentioned before, previous experiments relied on point by point comparisons of Fourier spectra. In order to have a more comprehensive evaluation of image statistics, we wrote a new IDL program that allows us to examine key statistics from our Fourier analysis for an entire image all at once. This new program takes the DFT of every pixel in a given data series and plots the results of several measurements graphically. Figure 5 shows the results of this test for frames 700-899 (20-seconds worth) of the 1999 data. The program then plots the power of the maximum peak both as a percentage of total signal and in sigmas (standard deviations) above the noise. It also plots the mean power and standard deviation, allowing us to look broadly at the statistics for an entire image all at once. In addition to its utility as a tool for analysis of the real data, we used this method to examine several of the models to which we now turn our attention.

Three questions remain for us to answer about our analysis. First, we must establish confidence intervals for peaks of particular heights. That is, we must establish how often peaks of $3\sigma$ and $4\sigma$ above the mean occur randomly as a result of noise in the data. Second, we must test our sensitivity to the kinds of complex oscillations that we would expect to see occurring naturally in the corona. A single pixel covers a significant area of a coronal loop, thus we expect to see oscillations that contain different phase components and varying frequencies, and are damped. If we can detect only simple oscillations, but are not sensitive to more realistic ones, the experiment cannot succeed. Finally, we must determine the role that alignment error plays in this experiment. Particularly, we want to answer the question of where it introduces noise, on the loops or elsewhere.

In order to establish confidence intervals, we used a 'Monte Carlo method.' At the heart of this method is the no-effect model described before, with one change. In addition to adding realistic image noise, we also applied random shifts of realistic magnitude. Once the model was run, the images were aligned, as described before, and Fourier transforms of 5 points were taken and smoothed with a 3 pixel boxcar. Statistics for each point were recorded and the program 'rolled the dice' again, that is, assembled a new artificial sequence and repeated the process. This was done a total of 625 times.

As we have already mentioned, we want to establish how often at least one peak above $3\sigma$ and $4\sigma$ occurred for each point. For the five points used, a $3\sigma$ peak occurred 22% of the time, while a $4\sigma$ peak occurred only about 1% of the time. Repeated runs using this method returned similar statistics. We thus consider $4\sigma$ as our discrimination factor.

We tested our sensitivity to complex oscillations using a system based on our no-effect model. We generated a 200-frame, no-effect sequence and inserted, into each pixel of a 20×20 region of bright corona, 4 separate damped oscillations each with randomly selected amplitudes (centered at 1% of pixel intensity), frequencies (centered at 1.25 Hz), and decay times, and a random phase factor. We then inserted noise and shifts as described above. The sequence was aligned and analyzed using our standard software package. Figure 6 shows the results of one such test, a map of the intensity of the maximum peak in the Fourier spectrum in sigmas (standard deviations). Strong peaks in the 20×20 region clearly stand out above the surrounding quiet pixels. An analysis of the statistics from several of these tests with these artificial data showed that on average 40% of pixels with non-uniform signal showed peaks above 3σ and 20% of pixels showed peaks above 4σ. These clearly exceed the established confidence intervals, confirming our sensitivity to oscillations of at least 1%.

We are left, finally, with the question of how image misalignment introduces noise into the data. Our result in the previous test is somewhat suggestive. While noise levels are increased due to image shifting throughout the data, areas where we did not insert signal have significantly lower noise levels (as we saw in Figure 6). However, the question that we especially want to answer is where exactly in the image noise is introduced in the data. If noise is introduced directly into the bright loops, it would hamper our ability to search for oscillations. In order to answer this question we return to our no-effect model, with a slightly different goal, discussed completely in Seaton (2001). This time, instead of a general summary of statistics, we look specifically for areas where noise levels are significantly increased. The top left panel of Figure 7 shows areas where, in the no-effect model, a peak of at least 3.5σ was detected (white areas). The upper right shows the actual data image after an unsharp-mask has been applied. This enhances the appearance of sharp edges and bright features. The bottom right panel of Figure 7 shows the two other panels overlaid on one another (pixels containing 3.5σ peaks are silhouetted in black against the lower corona). The results of this test are encouraging. As we expected, sharp gradients in the data do introduce some noise into the Fourier spectra. However, the bright loops where we search for oscillations are not contaminated.

With these tests complete, we can now compare the results of our simulations to the analysis of our data. The results shown, referring back to Figure 5, provide the best statistics for this analysis, since they occur near the end of the eclipse but before any saturation caused when the bright lower corona emission appeared.

Over a 4400 pixel region spanning almost the entire on-band half of the image, 29% of the pixels had a peak above 3σ for transforms of frames 700-899. While this percentage is slightly higher than the levels we got in the artificial no-effect models, it does not seem conclusive on its own. However, over a 1300 pixel region of the part of the data images with the brightest loops, 41% of the pixels had peaks above 3σ. And over an 1100 pixel region from a portion of the image with few bright loops, only 21% of the pixels had peaks above 3σ. Further, these relative statistics match the predictions of our data model. This comparison of real data with our artificial model suggests that we have detected oscillations where bright loops are present.

While we have already shown that image misalignment does not introduce noise directly into loops, we can check our data further to limit this possibility. Figure 8 shows four maps, of statistics from the power spectra taken of frames 700 through 899. Panel A shows a map of the frequency of the highest peak, B shows a map of where the highest peak fell between 0.75-1.0 Hz, C shows where the highest peak fell between 1.0-1.25 Hz, and D where the highest peak fell between 1.25-2.25 Hz. If all the power we detect came from shifting, we would expect to see the peaks at the same frequencies, even for loops with different physical parameters. While most of the power is concentrated in a narrow range from 0.75-1.0 Hz, there is enough loop-to-loop variation (especially in the brightest, on-loop pixels) to conclude that the enhanced power is not likely the result of image misalignment.

## 4. Conclusions

In this paper we have presented a new data set of the search for high-frequency coronal oscillations taken under near-perfect conditions at the eclipse of 11 August 1999, representing a significant improvement over our past observations. This improvement was the result of instrumental and environmental factors and because of the relatively brighter solar maximum corona at that eclipse. We have analyzed this data set using improved techniques, both for taking Fourier spectra of the data and for modeling artificial data sets that we use for comparison. These latter models answered several important questions about our data. First, a Monte Carlo approach revealed the confidence intervals for peaks that must be exceeded in order to confirm the presence of enhanced power as the result of coronal oscillations. Second, they suggested that the role of image misalignment in increasing noise is confined largely to loop edges, implying that Fourier peaks corresponding to bright, on-loop pixels are not likely to be the result of image misalignment. Finally, they revealed that our experiment should remain sensitive to the complex, quasi-periodic oscillations we expect to see in the data.

Comparing the results of these models against our own eclipse measurements, we find that our data show signs of increased power, particularly in areas where bright loops are present. Enhanced power is most present in the 0.75-1.0 Hz range, and is on of the order of 1% variation in total intensity.

Unfortunately, attempts to collect more oscillation data at the total eclipse of 21 June 2001 in Lusaka, Zambia, were foiled by a computer problem. Because wavelet analyses are more sensitive than Fourier analyses to finite-length wavepackets, we plan to supplement our Fourier analysis of future data with a wavelet analysis. Preparations to observe the eclipse of 4 December 2002 are under way.

## Acknowledgments

The work discussed in this paper was supported by grants from the National Science Foundation's Atmospheric Sciences Division (ATM-9707907; ATM-9812408; ATM-0000575), Astronomy Division (AST-9512216), and Education Division (DUE-9351279); the National Geographic Society through their Committee for Research and Exploration (grants: 5190-94; 5977-97; 6449-99; 6989-01); the Keck Northeast Astronomy Consortium; and NASA in the Guest Investigator Program for SOHO (EIT): NRA-98-03-SEC-051; and Williams College's Science Center and Safford Fund. Further support, in preparation, on-site, and during the analysis phase of this experiment by Stephan Martin, Jonathan Kern, Lee Hawkins, Misa Cowee, Mark Kirby, Timothy McConnochie, and Gabriel Brammer was invaluable. We are grateful to Magda Stavinschi and her staff at the Astronomical Institute of the Romanian Academy of Sciences for their assistance in arranging our observing site. We thank OceanOptics, Inc., educational products division for a grant and Inmarsat for the use of a satellite telephone. We also thank Harvey Tananbaum, Leon Golub, and Ed Deluca for their hospitality at the Harvard-Smithsonian Center for Astrophysics.

References

Aschwanden, M. J., Poland, A. I., and Rabin, D. M.: 2001, *Annu. Rev. Astron. Astrophys.* **39**: 175.

Berghmans, D., de Bruyne, P., and Goossens, M.: 1996, *Astrophys. J.* **472**, 398.

Berghmans, D., McKenzie, D., and Clette, F.: 2001, *Astron. Astrophys.* **369**, 291.

Cowsik, R., Singh, J., Saxena, A. K., Srinivasan, R., Raveendran, A. V.: 1997, *Solar Phys.* **188**, 89.

Golub, L. and Pasachoff, J. M.: 1997, *The Solar Corona*, Cambridge University Press, Cambridge, UK.

Golub, L. and Pasachoff, J. M.: 2001, *Nearest Star*, Harvard University Press, Cambridge, MA.

Gómez, D. O., Dmitruk, P. A., and Milano, L. J.: 2000, *Solar Phys.* **195**, 299.

Hollweg, J. V.: 1981, *Solar Phys.* **70**, 25.

Hollweg, J. V.: 1991, in P. Ulmschneider, E. R. Priest, and R. Rosner (eds.), *Mechanisms of Chromospheric and Coronal Heating*, Springer Verlag, Berlin, Germany, p. 423.

Ionson, J, A.: 1978, *Astrophys. J.* **226**, 650.

Ionson J. A.: 1979, private commmunication.

Krucker, S. and Benz, A. O.: 2000, *Solar Phys.* **191**, 241.

Krucker, S., Benz, A. O., Bastian, T. S., and Acton, L. W.: 1997, *Astrophys. J.* **488**, 499.

Pasachoff, J. M. and Ladd, E. F.: 1987, *Solar Phys.* **109**, 365.

Pasachoff, J. M. and Landman, D. A.: 1984, *Solar Phys.* **90**, 325.

Pasachoff, J. M., Babcock, B. A., Russell, K. D., McConnochie, T. H., and Diaz, J. S.: 2000, *Solar Phys.* **195**, 281.

Offman, L., Davila, J. M., and Shimizu, T.: 1996, *Astrophys. J. (Lett.)* **459**, L39.

Phillips, K. J. H., Read, P. D., Gallagher, P. T., Keenan, F. P., Rudawy, P., Rompolt, B., Berlicki, A., Buczylko, A., Diego, F., Barnsley, R., Smartt, R. N., Pasachoff, J. M., and Babcock, B. A.: 2000, *Solar Phys.* **193**, 259.

Porter, L. J., Klimchuk, J. A., and Sturrock, P. A.: 1994a, *Astrophys. J.* **435**, 482.

Porter, L. J., Klimchuk, J. A., and Sturrock, P. A.: 1994b, *Astrophys. J.* **435**, 502.

Priest, E. and Forbes, T.: 2000, *Magnetic Reconnection: MHD Theory and Applications*, Cambridge University Press, Cambridge.

Roberts, B.: 2000, *Solar Phys.*, **193** 139.

Russell, K. D.: 2000, *A Search for High-Frequency Coronal Oscillations at the 1999 Total Solar Eclipse*, Senior Honors Thesis, Williams College, Williamstown, Massachusetts.

Seaton, D. B.: 2001, *Continuing Science from the 1998 and 1999 Eclipses: High-Frequency Oscillations, General Structure, and Correlation with Space-Based Observations*, Senior Honors Thesis, Williams College, Williamstown, Massachusetts.

Shimizu, T: 1995, *PASJ* **47**, 251.

Singh, J., Cowsik, R., Raveendran, A. V., Bagare, S. P., Saxena, A. K., Sundararaman, K., Krishan, V., Naidu, N., Samson, J. P. A., and Gabriel, F.: 1997, *Solar Phys.* **170**, 235.

Williams, D. R., Phillips, K. J. H, Rudawy, P., Mathioudakis, M., Gallagher, P. T., O'Shea, E., Keenan, F. P., Read, P., and Rompolt, B.: 2001, *MNRAS* **326**, 428.

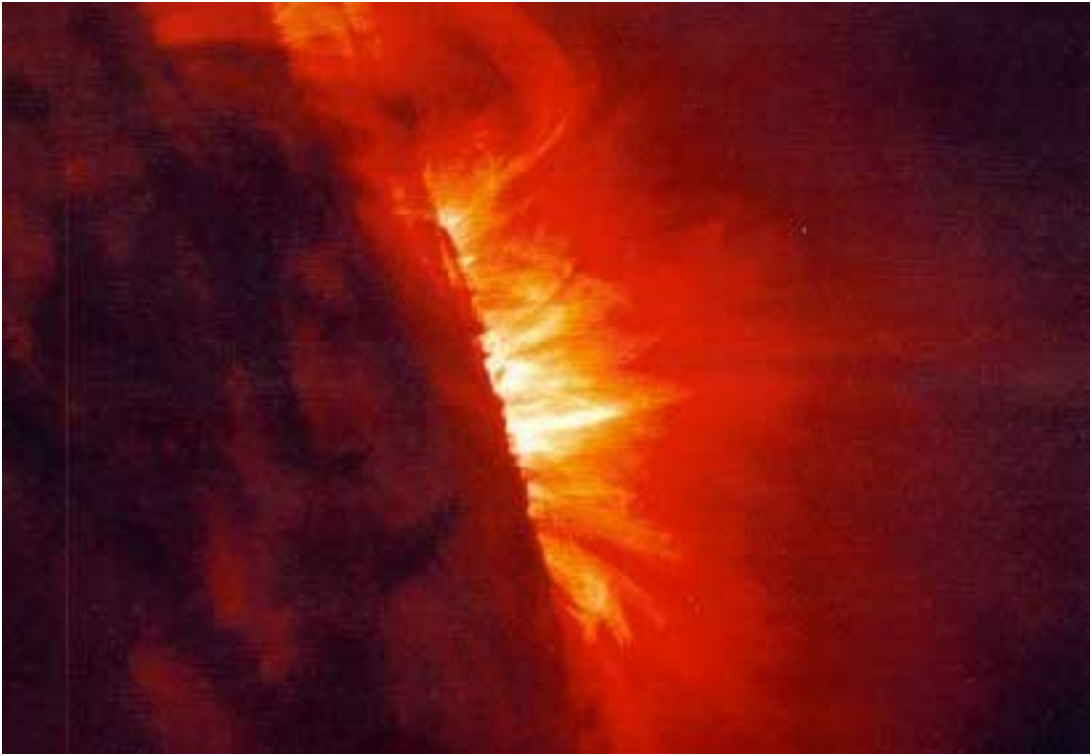

**Figure 1.** TRACE 195 Å image (Fe XII) showing loops observed at the 1999 eclipse.

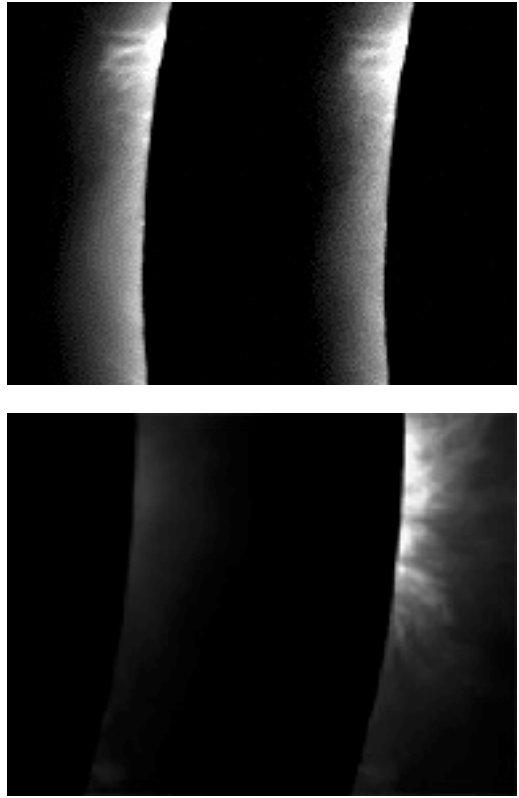

**Figure 2.** Data frames from 1998 (top) and 1999 (bottom). Both signal strength and clarity are improved in the 1999 data. For each year's data we see the image as it appeared on the CCD chip, with the continuum channel on the left and line channel on the right. The curved lunar limb with the inner solar corona is visible on each half of the chip.

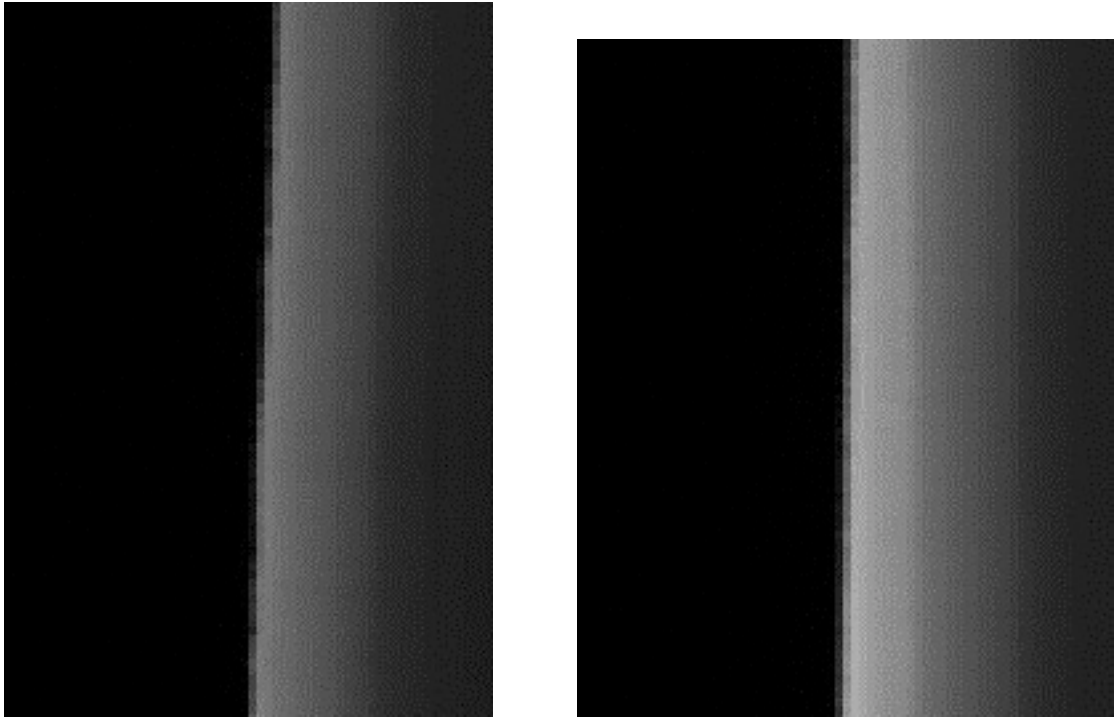

**Figure 3.** One row vs. time (vertical axis) before alignment (left) and after alignment (right). The dark lunar limb is stationary with respect to this row, thus, when aligned, the limb appears as the black vertical straight edge.

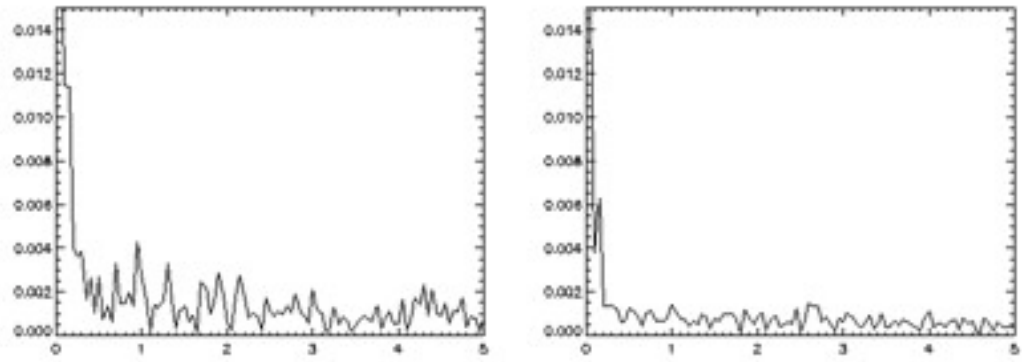

**Figure 4.** Fourier spectra of two neighboring pixels in the real data set, one on a bright loop (left) and one just off the same loop (right), revealing the localized nature of power in the spectra. Horizontal axis is in Hz, while the vertical axis is fraction of total power.

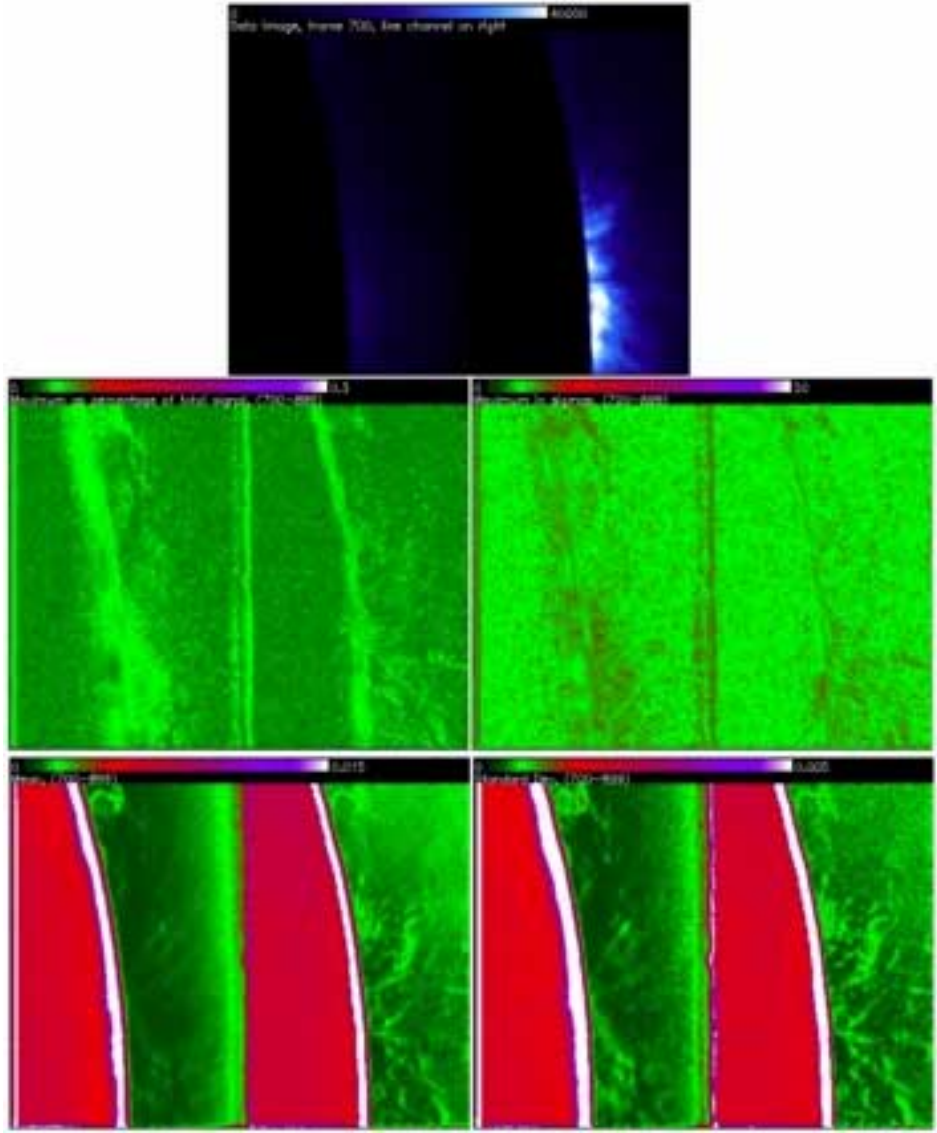

**Figure 5.** Maps of statistics from Fourier spectra for frames 700-899, for each image continuum channel is at left, line channel on the right. At top is a sample data image (frame 700). Center left shows the intensity of the maximum Fourier peaks for each pixel as a percentage of total signal while center right shows the maximum Fourier peak in sigmas. The bottom left is the mean power in the Fourier spectrum; bottom right is the standard deviation in the spectrum. It is not difficult to see how structure in the images corresponds to bright structures in the corona, as we would expect if we are detecting oscillations. The authors suggest viewing this figure in color on the accompanying CD-ROM, as a black-and-white reproduction cannot reveal the image scaling, which is displayed as a horizontal band at the top of each frame.

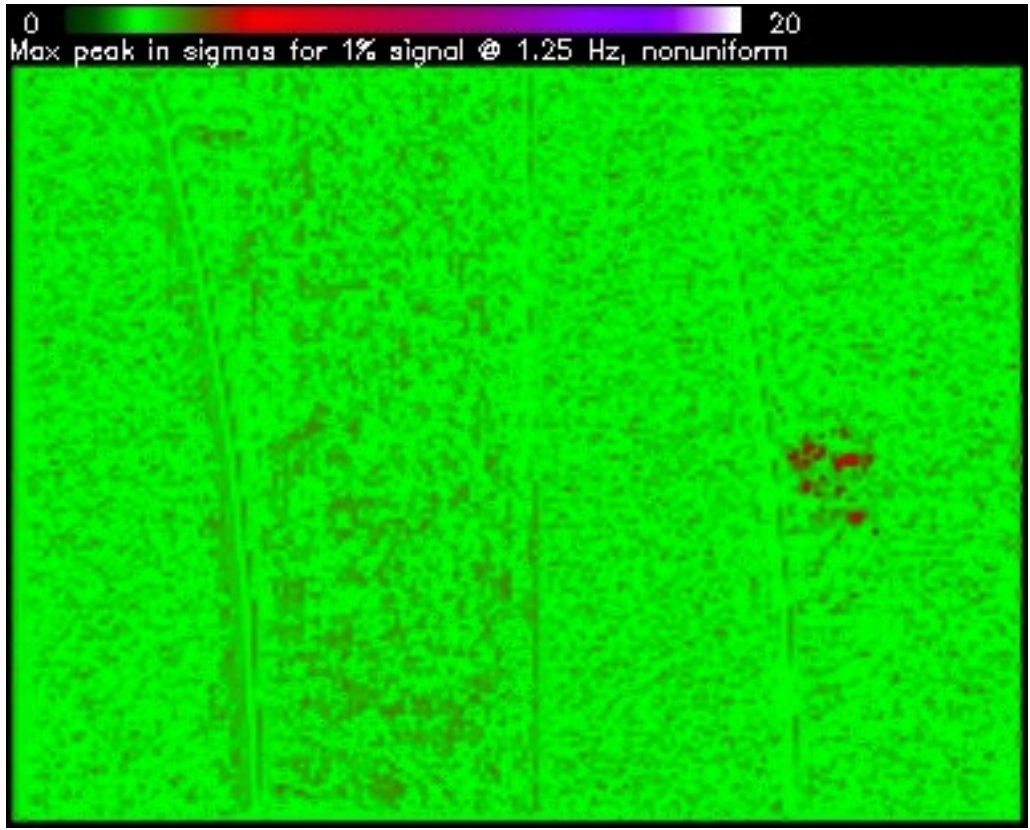

**Figure 6.** Results for one run with our complex signal model into which an artificial signal has been introduced showing the sensitivity of our method. Power is enhanced significantly in the 20×20 region of bright corona where signal was inserted into the line channel, which shows up as the blotchy area at center right. This model used 4 separate damped oscillations each with randomly selected amplitudes (centered at 1% of pixel intensity), frequencies (centered at 1.25 Hz), and decay times, and a random phase factor. The continuum channel, the left half of this image, was not considered in this test. The authors encourage the reader to view this figure in color.

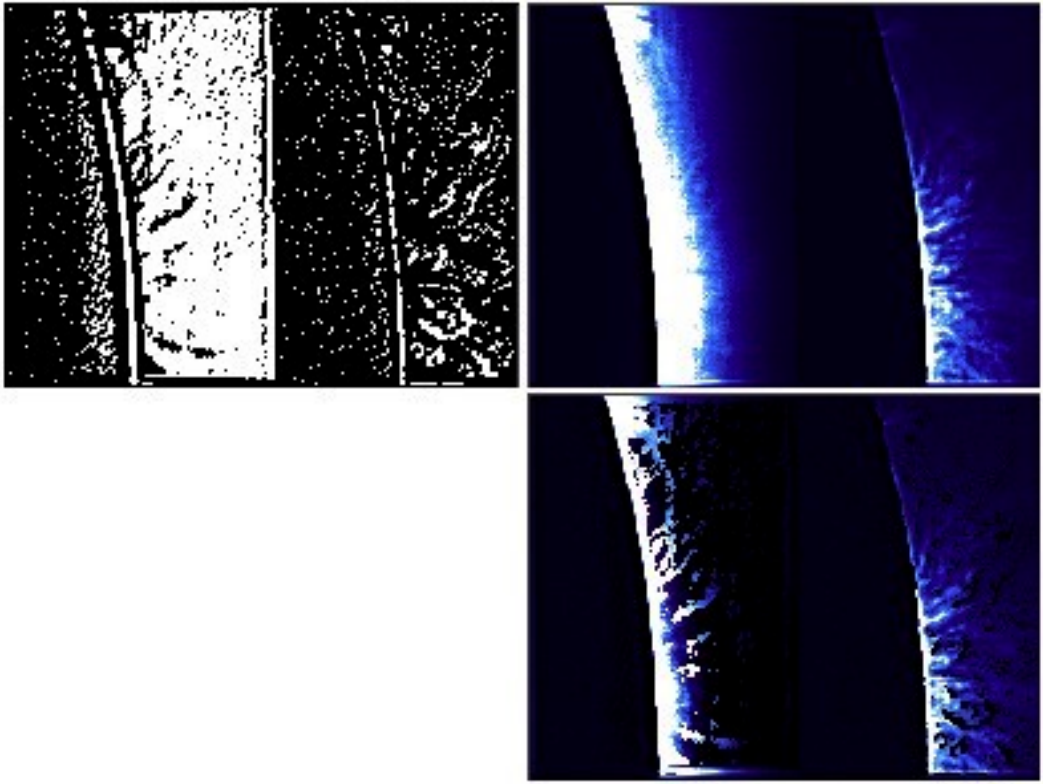

**Figure 7.** Map of results of no-effect model after aligning the artificial data showing where the strongest shift-related noise is induced in the image. As always in our displays, the continuum is at the left and on-band green-line image is at the right; we consider in this test only the on-band image. Upper left shows the image areas where the maximum Fourier peak was at least 3.5 sigmas (white areas). The upper right shows an unsharp-masked version of data frame 700. In the lower right we see the bright peaks overlain (black areas) on the unsharp-masked data image, revealing that, while shift-related noise is induced at loop edges, the bright loops themselves are noise free since the loops themselves are not negated by the noise from the upper left.. The continuum channel, on the left of each image, is disregarded in this test. The authors encourage the reader to view this figure in color, to see the results with highest level of contrast.

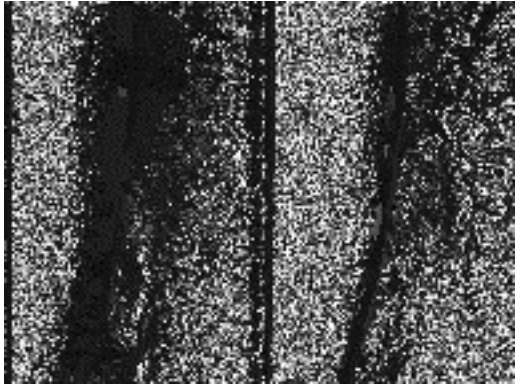 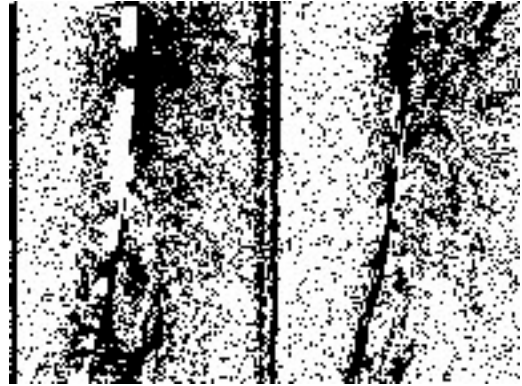

(a) Map of max peak frequency (darker means lower).   (b) Map of max peak frequencies from 0.75-1.0 Hz.

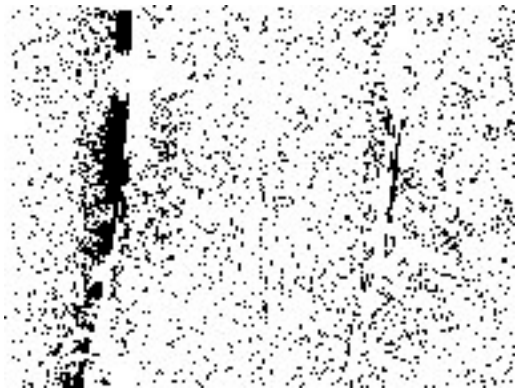 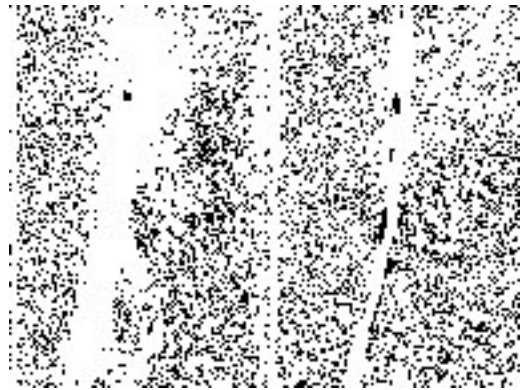

(c) Map of max peak frequencies from 1.0-1.25 Hz.   (d) Map of max peak frequencies from 1.25-2.25 Hz.

**Figure 8.** Maps showing the frequency at which the highest peak in the transform occurred for each pixel. (a) shows a map of the frequency of the maximum peak in the Fourier spectrum for each pixel—darker points had a lower maximum peak. (b) reveals pixels for which the maximum peak fell between 0.75-1.0 Hz, (c) pixels with maximum peaks between 1.0-1.25 Hz, and (d) pixels with maximum peaks between 1.25-2.25 Hz. By far the region of greatest enhanced power is the lowest frequency band.